\documentclass[a4paper,12pt]{article}

\usepackage[usenames]{color}
\usepackage{graphicx}
\usepackage{cite}
\usepackage{setspace}
\usepackage{bbm}
\usepackage{amsmath,amssymb,mathrsfs}

\newcommand{\im}{{\mathbbm i}}

\newcommand{\tr}{{\mbox{tr}}}

\begin{document}
\begin{titlepage}
\null
\begin{flushright}
August, 2017
\end{flushright}

\vskip 1.8cm
\begin{center}

  {\Large \bf Gauge Five-brane Solutions of Co-dimension Two}
\vskip 0.3cm 
 {\Large \bf in Heterotic Supergravity}

\vskip 1.8cm
\normalsize

  {\bf Shin Sasaki\footnote{shin-s(at)kitasato-u.ac.jp} 
and Masaya Yata\footnote{myata(at)ibs.re.kr}}

\vskip 0.5cm

  { \it
  ${}^1$Department of Physics \\
  Kitasato University \\
  Sagamihara 252-0373, Japan \\
\vspace{0.5cm}
  ${}^2$Center for Theoretical Physics of the Universe \\
  Institute for Basic Science (IBS) \\
   Seoul 08826, Republic of Korea
  }

\vskip 2cm

\begin{abstract}
We continue to study the BPS gauge five-brane solutions of codimension two in
ten-dimensional heterotic supergravity.
The geometry including the dilaton and the NS-NS $B$-field are sourced
 from the monopole chain in $\mathbb{R}^2 \times S^1$.
We find that the geometry is asymptotically Ricci flat and the dilaton is no longer imaginary valued. 
These properties are contrasted with the smeared counterpart discussed in our previous paper.
We perform the T-duality transformations of the solution and find that
 it never results in a non-geometric object.
\end{abstract}
\end{center}
\end{titlepage}

\newpage
\setcounter{footnote}{0}

\section{Introduction}
Extended objects known as branes play important roles in string
theories.
They have been utilized for studying supersymmetric gauge theories
\cite{Giveon:1998sr}, the AdS/CFT correspondence \cite{Maldacena:1997re}
and model buildings for particle physics in string theories
\cite{Blumenhagen:2006ci}.
In particular, the U-duality \cite{Hull:1994ys} is 
an important nature to understand the whole picture of string theories.
When the eleven-dimensional M-theory is compactified on $T^d$, there
appears the U-duality symmetry group $E_{d(d)}(\mathbb{R})$ in lower dimensions.
BPS branes which preserve fractions of supersymmetry 
in lower dimensions are classified according to the U-duality symmetry group
$E_{d(d)}(\mathbb{R})$. 
The higher dimensional origin of these lower-dimensional BPS branes 
are wrapped/unwrapped F-strings, D-branes, NS5-branes, Kaluza-Klein (KK)
5-branes and waves.
There are exceptions, however, in lower than eight-dimensional
space-time. BPS states whose origin cannot be traced back to
these ordinary branes appear in the $E_{d(d)}(\mathbb{R})$ U-duality
multiplet \cite{Elitzur:1997zn,Obers:1998fb, Blau:1997du}.
These BPS states are called exotic states whose higher dimensional
origin are known as exotic branes
\cite{Eyras:1999at,LozanoTellechea:2000mc}.

Among other things, an exotic brane called $5^2_2$-brane in type II
string theories have been intensively studied 
\cite{Kikuchi:2012za,Kimura:2013fda, 
Andriot:2014uda,Kimura:2014upa, Chatzistavrakidis:2013jqa, Kimura:2015yla,
Okada:2014wma, Sakatani:2014hba, Sakatani:2016sko,Lee:2016qwn}.
The exotic $5^2_2$-brane is a solitonic five-brane of codimension two,
whose tension is proportional to $g_{s}^{-2}$, as its name stands for \cite{Obers:1998fb}.
The $5^2_2$-brane has the $U(1)^2$ isometry along the transverse directions to the
brane worldvolume. 
It is a defect brane \cite{Okada:2014wma, Bergshoeff:2011se} and has a number of specific properties.
For example, exotic branes are non-geometric objects
\cite{deBoer:2010ud,Hull:2004in, Hassler:2013wsa}, namely, 
the background metric and other supergravity fields for an exotic brane
are governed by multi-valued functions of space-time.
However, monodromies associated with the exotic branes are indeed given
by the U-duality group. This means that the geometry is patched together
by the symmetry and 
it is completely a consistent solution in string theories. 
In this sense, they are not described by ordinary manifolds but their generalizations, called U-folds.
The $5^2_2$-branes in type II supergravities 
are obtained through the chain of T-duality transformations of
the NS5-branes. The monodromy of the $5^2_2$-brane geometry is given by
the $O(2,2)$ T-duality group. Therefore it is an explicit example of the T-fold.
Some efforts has been devoted to understand the exotic branes in
type II string theories. 
However, exotic branes in heterotic and type I string theories are
poorly understood. 

In our previous paper \cite{Sasaki:2016hpp}, we have studied the
T-duality chain of five-branes in heterotic supergravity.
It is known that there are three distinct five-branes in heterotic
theories \cite{Callan:1991dj, Strominger:1990et,Duff:1990wv,Rey:1991uu}.
They are called the symmetric, neutral and gauge types 
\cite{Callan:1991dj, Strominger:1990et}.
In order to perform the T-duality transformations of these five-branes,
we have introduced $U(1)^2$ isometry along the transverse directions to
the branes by the smearing procedure.
Since the heterotic supergravity action contains the kinetic term of the
non-Abelian gauge field and the $R^2$ term which are the leading
order in the $\alpha'$ corrections, the well-known Buscher rule
\cite{Buscher:1987sk} is modified in heterotic theories \cite{Tseytlin:1991wr,
Bergshoeff:1995cg, Serone:2005ge}. 
Accordingly, the generalized metric is also modified from the one in type II theories.
Using the modified Buscher rule in heterotic theories we have derived
the KK5-branes and $5^2_2$-branes of the three kinds.
We found that the monodromies of the $5^2_2$-branes for the symmetric and the neutral types
are given by the $O(2,2)$ T-duality group and they are therefore 
non-geometric objects.
On the other hand, we were faced with some difficulties for the gauge
type five-brane.
We found that the smearing procedure makes the metric of the gauge type solution
ill-defined. A function that governs the solution becomes
negative valued at some regions in space-time and the dilaton $\phi$ becomes
imaginary valued in there.
In order to make the gauge type solution of codimension two
well-defined, we need to abandon the smearing procedure and re-consider the
gauge five-brane from the first principle.

In this paper we make a complementary study of the gauge five-brane of
codimension two in heterotic supergravity.
We will re-construct the gauge five-brane based on the well-behaved
monopole solution of codimension two which is known as a monopole chain
\cite{Ward:2005nn}. We will show that the geometry is asymptotically
Ricci flat and the solution is well-behaved except the origin of the brane position.
We will perform the chain of the T-duality transformations on the
solution and find that the $5^2_2$-brane obtained through the new gauge
five-brane is not a non-geometric object.
The organization of this paper is as follows.
In the next section, we introduce the gauge five-brane solution in
heterotic supergravity.
In section 3, we introduce the Nahm construction of monopole in
$\mathbb{R}^2 \times S^1$ in the large $S^1$ limit.
In section 4, we write down the gauge five-brane solution of codimension
two.
In section 5, we analyze the T-duality transformations of the gauge
five-brane. The background supergravity fields for the gauge KK5-brane and
$5^2_2$-brane are obtained.
Section 6 is conclusion and discussions.
A version of the KK5-brane in another T-duality route is found in
Appendix A.

\section{Five-brane solutions in heterotic supergravity}
The low-energy effective theory of heterotic string theory is the
ten-dimensional heterotic supergravity.
The action in the $\mathcal{O} (\alpha')$ is given by \cite{Bergshoeff:1988nn}
\begin{align}
S =& \ \frac{1}{2 \kappa^2_{10}} \int \! d^{10} x \sqrt{-g} e^{-2\phi} 
\Bigl[
R (\omega) - \frac{1}{3} \hat{H}_{MNP}^{(3)} \hat{H}^{(3)MNP} + 4 \partial_M \phi \partial^M
 \phi 
\notag \\
& 
\qquad 
\qquad 
\qquad 
\qquad 
\qquad 
 \frac{}{}
+ \alpha' 
\left(
\text{Tr} F_{MN} F^{MN} + R_{MN AB} (\omega_{+}) R^{MNAB} (\omega_{+})
\right)
\Bigr].
\label{eq:heterotic_sugra}
\end{align}
Here we have used the convention 
$\frac{\kappa^2_{10}}{2g^2_{10}} =\alpha'$ where
 $\kappa_{10}$ and $g_{10}$ are the gravitational
and the gauge coupling constants in ten dimensions.
The dilaton is denoted as $\phi$.
The metric $g_{MN} \ (M,N=0, \ldots, 9)$ is defined
by the vielbein as $g_{MN} = \eta_{AB} e_M {}^A e_N {}^B$.
Here $A,B = 0, \ldots, 9$ are indices in the local Lorentz frame 
and we employ the mostly plus convention of the flat metric $\eta_{AB} =
\text{diag} (-1, 1, \ldots, 1)$.
The local Lorentz indices $A, B, \ldots$ are lowered and raised by
$\eta_{AB}$ and its inverse $\eta^{AB}$.
The Ricci scalar $R (\omega)$ and the Riemann tensor $R^{AB} {}_{MN}$ are defined by the spin connection
$\omega_{M} {}^{AB}$ as 
\begin{align}
R (\omega) =& \ e^M {}_A e^N {}_B R^{AB} {}_{MN} (\omega), \notag \\
R^{AB} {}_{MN} (\omega) =& \ \partial_M \omega_N {}^{AB} - \partial_N
 \omega_M {}^{AB} + \omega_M {}^{AD} \eta_{CD} \omega_N {}^{CB} -
 \omega_N {}^{AD} \eta_{CD} \omega_M {}^{CB}.
\label{eq:Ricci}
\end{align}
The spin connection is expressed by the vielbein $e_M {}^A$ and its
inverse:
\begin{align}
\omega_{M}^{~AB}={1\over2} 
\Bigl[
e^{AN}( \partial_{M}e_{N}^{~B} - \partial_{N}e_{M}^{~B} )
-e^{BN}( \partial_{M}e_{N}^{~A} - \partial_{N}e_{M}^{~A} )
-e^{AP} e^{BQ}
( \partial_{P} e_{Q C} - \partial_{Q}e_{P C} ) 
e_{M}^{~C}
\Bigr].
\end{align}
The Riemann tensor in the $\mathcal{O} (\alpha')$ action
\eqref{eq:heterotic_sugra} is defined through the modified spin
connection $\omega_{\pm M} {}^{AB}$.
This is defined by 
\begin{align}
\omega_{\pm M} {}^{AB} = \omega_{M} {}^{AB} \pm \hat{H}_{M}^{(3)} {}^{AB},
\label{eq:modified_spin_connection}
\end{align}
where $\hat{H}_M^{(3)} {}^{AB} = e^{NA} e^{PB} \hat{H}_{MNP}^{(3)}$ is
the modified $H$-flux.
The modified flux $\hat{H}_{MNP}^{(3)}$ is defined by 
\begin{align}
\hat{H}^{(3)}_{MNP} =& \ H_{MNP}^{(3)} + \alpha' 
\left(
\Omega_{MNP}^{\text{YM}}
- 
\Omega_{MNP}^{\text{L}+}
\right)
+ \mathcal{O} (\alpha^{\prime 2}).
\label{eq:H-flux}
\end{align}
Here $H_{MNP}^{(3)}$ is the ordinary field strength of the NS-NS $B$-field:
\begin{align}
H_{MNP}^{(3)} =& \ \frac{1}{2} (\partial_M B_{NP} + \partial_N B_{PM} +
 \partial_P B_{MN}).
\end{align}
The Yang-Mills and the Lorentz Chern-Simons terms in \eqref{eq:H-flux} are defined by 
\begin{align}
\Omega^{\text{YM}}_{MNP} =& \ 
3! \text{Tr}
\left(
A_{[M} \partial_N A_{P]} + \frac{2}{3} A_{[M} A_{N} A_{P]}
\right), \notag \\
\Omega^{\text{L}+}_{MNP} =& \ 
3!
\left(
\eta_{BC} \eta_{AD}\omega_{+[M} {}^{AB} \partial_{N} \omega_{+P]} {}^{CD} 
+ \frac{2}{3} \eta_{AG} \eta_{BC} \eta_{DF} \omega_{+[M}^{AB} \omega_{+N} {}^{CD}
 \omega_{+P]} {}^{FG}
\right).
\end{align}
Here $A_M = A^I_M T^I$ is the Yang-Mills gauge field and 
$T^I \ (I,J,K = 1, \ldots, \text{dim} \mathcal{G})$ are 
the generators of the Lie algebra 
$\mathcal{G}$ associated with the gauge group $G$.
The gauge group $G$ is $SO(32)$ or $E_8 \times E_8$ depending on the
heterotic string theories we consider. 
The symbol $[M_1 M_2 \cdots M_n]$ stands for the anti-symmetrization of
indices with weight $\frac{1}{n!}$. 
The modified $H$-flux $\hat{H}^{(3)}$ obeys the Bianchi identity:
\begin{align}
d \hat{H}^{(3)} = \alpha'
\left(
\text{Tr} F \wedge F - \text{Tr} R \wedge R
\right) + \mathcal{O} (\alpha^{\prime 2}),
\label{eq:Bianchi}
\end{align}
where $R^{AB} = \frac{1}{2!} R^{AB} {}_{MN} d x^M \wedge d x^N$ is the
$SO(1,9)$-valued curvature 2-form.
The component of the Yang-Mills gauge field strength 2-form 
$F = \frac{1}{2!} F_{MN} d x^M \wedge dx^N$ is given by 
\begin{align}
F_{MN} {}^I = \partial_M A_N^I - \partial_N A_M^I + f^I {}_{JK} A^J
 {}_M A^K {}_N,
\end{align}
where $f^I {}_{JK}$ is the structure constant for $\mathcal{G}$.

The 1/2 BPS ansatz for the five-brane solution is given by
\cite{Strominger:1990et, Callan:1991dj}
\begin{align}
& \ ds^2 = \eta_{\mu\nu} dx^{\mu} dx^{\nu} + H (x) \delta_{mn} dx^m dx^n, \notag \\
& \ \hat{H}^{(3)}_{mnp} = \mp \frac{1}{2} \varepsilon_{mnpq} \partial_q H (x), \qquad 
e^{2\phi} = H (x), \notag \\
& F_{mn} = \pm \tilde{F}_{mn} = \pm \frac{1}{2} \varepsilon_{mnpq} F^{pq}, 
\qquad A_{\mu} = 0,
\label{eq:five-branes}
\end{align}
where the indices $\mu,\nu = 0,5,6,7,8,9$ stand for the
world-volume while $m,n = 1,2,3,4$ represent the transverse directions to the five-branes.
The Levi-Civita symbol is denoted as  $\varepsilon_{mnpq}$.
The gauge field $A_m$ satisfies the (anti)self-duality condition in the
transverse four dimensions.
By using the Bianchi identity \eqref{eq:Bianchi} together with the ansatz
\eqref{eq:five-branes}, we find that the equation for the 
``$H$-function'' $H(x)$
reduces to
\begin{align}
\Box H = \pm \alpha' \mathrm{Tr} [F_{mn} \tilde{F}^{mn}] + \cdots,
\label{eq:source}
\end{align}
where 
$\Box$ is the Laplacian in the four-dimensional transverse space and 
$\cdots$ are terms involving the Riemann curvature.
The equation \eqref{eq:source} means that the source term in the right hand side of the Poisson
equation for 
$H(x)$
 is given by the Yang-Mills instanton density and the
Riemann curvatures.

There are three distinct solutions to the 1/2 BPS five-brane conditions
\eqref{eq:five-branes}. 
They are so called symmetric, neutral \cite{Callan:1991dj} and gauge
five-brane solutions \cite{Strominger:1990et}. 
The symmetric solution is an exact solution in $\alpha'$-expansion 
while the neutral and the gauge solutions are valid in $\mathcal{O} (\alpha')$.
For the neutral and the gauge solutions, the Riemann curvature terms in
\eqref{eq:source} become higher orders in $\alpha'$ expansion and are
neglected. 
The bulk gauge field becomes trivial in the neutral solution and it
is just the NS5-brane in type II supergravities.
On the other hand, the gauge solution involves the non-trivial bulk
Yang-Mills field configuration. 
Indeed, the gauge field is given by the instanton configuration with the 
non-zero topological charge $k$ defined by 
\begin{align}
k = - \frac{1}{ 32 \pi^2} \int \mathrm{Tr} [F \wedge F],
\end{align}
where the integral is defined in the transverse four-space.
Once a Yang-Mills instanton solution is obtained, 
the 
$H$-function $H (x)$ 
is determined through the relation \eqref{eq:source}.
The other quantity that distinguishes the five-branes 
is the charge $Q$ associated with the modified $H$-flux:
\begin{align}
Q = - \frac{1}{2\pi^2 \alpha'} \int_{S^3} \! \hat{H}^{(3)}.
\end{align}
Here $S^3$ is the asymptotic three-sphere surrounding the five-branes.
The symmetric, neutral and the gauge five-branes have charges $(k,Q) =
(1,n), (0,n), (1,8)$ respectively.

In the previous paper, we studied the T-duality transformations of
these heterotic five-brane solutions.
In order to perform the T-duality transformations, we introduced
$U(1)^2$ isometry along the transverse directions to the five-branes.
To this end, we employed the smearing method for which the
dimensionality of the space, where the 
$H$-function is defined, is effectively reduced \cite{Ortin:2015hya}.
Then we wrote down the explicit solutions of the codimension two
five-branes with the $U(1)^2$ isometry.
After performing the chain of T-dualities, we constructed the $5^2_2$-branes of three kinds.
We found that the symmetric and the neutral $5^2_2$-branes are T-fold
with the non-trivial $O(2,2)$ monodromy.
On the other hand, the gauge $5^2_2$-brane solution is ill-defined in the sense that
the dilaton becomes imaginary valued 
in some regions of space-time. 
This is due to the fact that the right hand side of the equation
\eqref{eq:source} that we have assumed is not appropriate one.

For the concreteness, we start from the gauge five-brane of codimension
four. 
A typical example of the self-dual solution is provided by  
the BPST one-instanton in the non-singular gauge \cite{Belavin:1975fg}.
By employing this configuration as the Yang-Mills field, 
the solution is given by \cite{Callan:1991dj}
\begin{align}
A_m = -\frac{ \sigma_{mn} x^n}{r^2 + \rho^2}, 
\qquad 
H (r) = e^{2\phi_0} + 8 \alpha' \frac{r^2 + 2 \rho^2}{(r^2 +
 \rho^2)^2},
\qquad 
B_{mn} = \Theta_{mn},~~~(\text{constant} ),
\label{eq:gauge_solution}
\end{align}
where 
$r^2 = (x^1)^2 + (x^2)^2 + (x^3)^2 + (x^4)^2$ and 
the gauge field takes value in the $SU(2)$ subgroup of $G$.
Here $\sigma_{mn}$ is the $SO(4)$ Lorentz generator and $\phi_0, \rho$ are constants.
In order to introduce the $U(1)$ isometries in the transverse
directions and reduce the codimension of the solution, we look for the
self-duality solution to the Yang-Mills gauge field in lower dimensions.
It is well known that the self-duality equation for the Yang-Mills gauge
field becomes that of monopoles of codimension three by the dimensional reduction.
Indeed, in \cite{Khuri:1992hk, Gauntlett:1992nn}, the gauge five-brane
solutions of codimension three based on the regular BPS monopoles was obtained.
In the previous paper \cite{Sasaki:2016hpp}, we constructed 
a gauge five-brane of codimension three by introducing the smeared
 instantons in $\mathbb{R}^3 \times S^1$ in the right hand side of \eqref{eq:source}.
This is a naive limit of the solution in
\cite{Khuri:1992hk,Gauntlett:1992nn} where the radius in $\mathbb{R}^3
\times S^1$ becomes small (see fig. \ref{T-duality} in section 6.).
Proceeding further in this way, we have constructed the gauge
five-brane solution of codimension two based on the smeared monopole in $\mathbb{R}^2
\times S^1$. 
We found that the $H$-function 
$H(\vec{x})$ associated with this
solution is given by \cite{Sasaki:2016hpp}
\begin{align}
H = e^{2\phi_0} - \frac{\alpha' \tilde{\sigma}^2}{2 r^2 
\left(
\tilde{h}_0 - \frac{\tilde{\sigma}}{2} \log (r/\mu)
\right)^2
}, \qquad 
r^2 = (x^1)^2 + (x^2)^2, 
\end{align}
where $\phi_0, \tilde{\sigma}, \tilde{h}_0, \mu$ are constants.
This is obviously not positive definite. 
As a result, the dilation becomes imaginary valued
at some points near the core of the brane.
This indicates the fact that the smeared monopole does not work as a
source of the well-defined brane geometry of codimension two.

In the following, we replace the right hand side of \eqref{eq:source}
with more appropriate solution, namely, the well-behaved periodic
monopole and examine the gauge five-brane solution of codimension two again.

\section{Nahm construction for monopoles of codimension two}
In this section, we introduce the monopole solution of codimension two
which will provide a well-defined brane geometry.
There is a systematic mathematical program to find analytic solutions of
monopoles, known as the Nahm construction \cite{Nahm:1979yw, Hitchin:1983ay}.
The monopole of codimension two that we consider in this paper is 
just the small $S^1$ limit of a periodic monopole defined in $\mathbb{R}^2 \times S^1$.
In the following, we write down the explicit field configuration 
that is based on the periodic monopole solution discussed in \cite{Ward:2005nn}.

The BPS monopole equation in $\mathbb{R}^3$ is defined as\footnote{In
the following we employ the minus sign in the right hand side. It is
possible, of course, to find the solution for the other sign.}
\begin{align}
D_i \Phi = - B_i, \quad (i = 1,2,3).
\label{eq:monopole}
\end{align}
Here $\Phi$ is an adjoint scalar field and $B_i$ is the magnetic field
defined through the gauge field $A_i$.
They belong to the adjoint representation of a gauge group $G$ with
an anti-hermitian matrix.
For definiteness, we consider the $G = SU(2)$ gauge group.
The relevant quantities are defined by 
\begin{align}
D_i \Phi = \partial_i \Phi + [A_i, \Phi], \quad 
F_{ij} = \partial_i A_j - \partial_j A_i + [A_i, A_j], \quad 
B_i = \frac{1}{2} \varepsilon_{ijk} F_{jk}.
\end{align}
We note that the equation \eqref{eq:monopole} is obtained via the
dimensional reduction of the (anti)self-duality equation $F_{mn} = -
\tilde{F}_{mn}$ in $\mathbb{R}^4$. 
The adjoint scalar field $\Phi$ is identified with the gauge field
component of the compact direction.

We now compactify one of the three-dimensional direction in $\mathbb{R}^3$ to $S^1$ and
consider the equation \eqref{eq:monopole} in $\mathbb{R}^2 \times S^1$.
We define the coordinate on $\mathbb{R}^2 \times S^1$ by $\vec{x} = (x^1,x^2,x^3)\equiv
(x,y,z)$ and the $S^1$ direction has the periodicity $z \sim z + \beta$.
Here $\beta = 2\pi R$ and $R$ is the radius of $S^1$.
We are looking for the solution to the equation \eqref{eq:monopole} in
the small-$\beta$ limit.
In this limit, the equation \eqref{eq:monopole} is effectively defined in two dimensions.
For the ordinary 't Hooft monopole of codimension three, the gauge group is broken down to
$U(1)$ at infinity and the asymptotic behavior of the 
solution is governed by the Abelian Dirac monopole.
We therefore employ the same boundary condition for our case.
By using the Bianchi identity, the Abelian reduction of the monopole
equation \eqref{eq:monopole} on $\mathbb{R}^2$ becomes
\begin{align}
\left(
\frac{\partial^2}{\partial x^2} + \frac{\partial^2}{\partial y^2} 
 \right)
\Phi = 0.
\end{align}
This is the Laplace equation in two dimensions whose 
spherically symmetric solution is given by 
\begin{align}
\Phi = c_1 \log r + c_2, \qquad r^2 = x^2 + y^2.
\label{eq:bc}
\end{align}
Here $c_1,c_2$ are constants.
This is the boundary condition of the $SU(2)$ monopoles in $\mathbb{R}^2
\times S^1$.
Cherkis and Kapustin claimed that the BPS monopoles defined in $\mathbb{R}^2
\times S^1$ are the Nahm dual to solutions to the Hitchin system in
$\mathbb{R} \times S^1$ \cite{Cherkis:2000cj}.
By the Nahm transformation, the solution to \eqref{eq:monopole} with 
the boundary condition \eqref{eq:bc} is given by 
\begin{align}
\Phi = \im \int^{\infty}_{-\infty} \! d u \int^{\pi/\beta}_{-\pi/\beta} \! dv \ u
 \Psi^{\dagger} \Psi, \qquad 
A_i = \int^{\infty}_{- \infty} \! d u \int^{\pi/\beta}_{-\pi/\beta} \! dv \ \Psi^{\dagger}
 \partial_i \Psi, \qquad (i=1,2,3).
\label{eq:Nahm_transformation}
\end{align}
Here $(u,v)$ are coordinates of the dual space $\mathbb{R} \times S^1$
and $2\pi/\beta$ is the dual period of $S^1$.
The ``Dirac zero-mode'' $\Psi = \Psi (u,v; \vec{x})$ has a $2 \times 2$ matrix
representation satisfying the following relation:
\begin{align}
\Delta \Psi = 0, \quad 
\int^{\infty}_{- \infty} \! du \int^{\pi/\beta}_{-\pi/\beta} \! dv \
 \Psi^{\dagger} \Psi = \mathbf{1}_2.
\label{eq:Hitchin}
\end{align}
Here $\Delta$ is the Dirac operator given by \cite{Cherkis:2000cj}
\begin{align}
& \ 
\Delta =
\left[
\begin{array}{cc}
2 \partial_{\bar{s}} - z & P(s) \\
P^{*} (\bar{s}) & 2 \partial_s + z
\end{array}
\right], \quad 
P(s) = C \cosh (\beta s) - \zeta, 
\notag \\
& \ 
s = u + \im v, \ u \in \mathbb{R}, \ v \in
 \left[ -\pi/\beta,\pi/\beta \right), 
\quad
\zeta = x + \im y, \quad z \sim z + \beta.
\label{eq:Dirac_operator}
\end{align}
Note that $x,y,z$ have the mass dimension $-1$ while the dual coordinates $s,u,v$
have the dimension $+1$. The function $P$ is determined by the periodic Hitchin fields.
A dimensionful constant $C$ is recognized as the
size of the monopole and it can be compared with the period $\beta$.
The small $C$ means $C \ll \beta$, namely, the decompactification limit $R \to \infty$.
In this limit, the Hitchin equation reduces to the Nahm equation for the monopole in
$\mathbb{R}^3$ \cite{Maldonado}.

We are interested in the solution in the large-$C$ (or equivalently small-$\beta$) limit. 
In this limit, the radius of the physical circle in $\mathbb{R}^2 \times
S^1$ becomes small $R \to 0$ and the monopoles exhibit the isometry
along $S^1$.
Now we solve the Dirac equation $\Delta \Psi = 0$.
To this end, we look for functions $f = f (u,v ; \vec{x}), g = g (u,v;
\vec{x})$ that satisfy
\begin{align}
\Delta 
\left[
\begin{array}{c}
g \\
f
\end{array}
\right] = 
\left[
\begin{array}{c}
2 g_{\bar{s}} - z g + P (s) f \\
2 f_s + z f + P^{*} (\bar{s}) g
\end{array}
\right]
=
0.
\label{eq:fg_eq}
\end{align}
As discussed in \cite{Ward:2005nn}, in the region where $P(s)$ is not zero,
the solution to the equation \eqref{eq:fg_eq} becomes trivial $f = g = 0$.
The exception is the points where $P (s) = 0$.
In order to find the non-trivial solution for $f,g$, we first define a
zero point of $P(s_0) = 0$, namely,
\begin{align}
s_0 = u_0 + \im v_0 = \frac{1}{\beta} \cosh^{-1} (\zeta/C),
\end{align}
or more explicitly
\begin{align}
u_0 =& \ \frac{1}{\beta} 
\cosh^{-1} \frac{1}{2C} (\sqrt{(C + x)^2 + y^2}
 + \sqrt{(C - x)^2 + y^2}), \notag \\
v_0 =& \ \frac{1}{\beta} \cos^{-1} \frac{1}{C} (\sqrt{(C + x)^2
 + y^2} - \sqrt{(C - x)^2 + y^2}) + n, \quad (n \in \mathbb{Z}).
\label{eq:rt_def}
\end{align}
Since $\text{cosh} (x)$ is an even function of $x$, the zeros are in
fact given by $s = \pm s_0$.
When these zeros are degenerate, $s_0 = - s_0$, we find $x = \pm C, y = 0$. 
On top of the zero $s=s_0$ we have $P = 0$ and the solutions to the
above equation are 
\begin{align}
g = \exp \left( \frac{z}{2} \bar{s} \right), 
\quad 
f = \exp \left( - \frac{z}{2} s \right).
\end{align}
When one leaves from the zero $s=s_0$, 
we have $f = g = 0$ as discussed.
Indeed, $P$ is a continuous function of $s$ and one can reach the point $s_0$ continuously.
Therefore $f,g$ are continuous functions of $s = u + \im v$ whose support
is localized around $s \simeq s_0$. In order to find a solution, we expand $P(s)$
around $s = s_0$ and find 
\begin{align}
P(s) = P(s_0) + \left. \frac{\partial P}{\partial s} \right|_{s =
 s_0} (s-s_0) + \cdots
\simeq \beta \xi (s-s_0).
\end{align}
Here we have defined $\xi (x,y) = C \sinh (\beta s_0)$.
Then, around the zero $s \sim s_0$, the equation \eqref{eq:fg_eq} becomes
\begin{align}
\begin{array}{l}
2 g_{\bar{s}} - z g + \beta \xi (s - s_0) f = 0, 
\\
2 f_s + z f + \beta \bar{\xi} (\bar{s} - \bar{s}_0) g = 0.
\end{array}
\label{eq:eqs_fg}
\end{align}
It is easy to confirm that $E (s-s_0)$ defined by the following function satisfies the
equations \eqref{eq:eqs_fg}:
\begin{align}
E (s) = \exp \left[- \frac{\beta}{2} |\xi| s \bar{s} - \frac{z}{2} (s - \bar{s}) \right].
\end{align}
This function $E(s-s_0)$ has a peak at $s \sim s_0$ and decays exponentially
outside the support.
Using this expression together with the fact that the zero points are
indeed $s= \pm s_0$, we have solutions to \eqref{eq:fg_eq}:
$f(s) = \frac{|\xi|}{\xi} E
(s \pm s_0)$, $g(s) = \pm E (s \pm s_0)$. 
Then, by using these functions, the solution to the Dirac equation is given by 
\begin{align}
& \Psi \simeq 
\sqrt{\frac{\beta}{2\pi}}
|\xi|^{-\frac{1}{2}} 
\left[
\begin{array}{cc}
\xi E (s-s_0) & - \xi E (s+s_0) \\
|\xi| E (s-s_0) & |\xi| E (s+s_0)
\end{array}
\right].
\label{eq:psi_sol}
\end{align}
Here we have introduced the overall factor for the normalization.
Indeed, one calculates
\begin{align}
\Psi^{\dagger} \Psi \simeq \ 
= \frac{\beta}{\pi} |\xi| 
\left[
\begin{array}{cc}
|E_{-}|^2 & 0 \\
0 & |E_{+}|^2
\end{array}
\right].
\end{align}
Here we have defined $E_{\pm} = E (s \pm s_0)$ and 
\begin{align}
|E_{\pm}|^2 = 
\exp 
\left[
- 2 \pi |\xi| (u \pm u_0)^2
\right]
\exp
\left[
- 2 \pi |\xi| (v \pm v_0)^2
\right].
\end{align}
The integration by $u$ in \eqref{eq:Hitchin} is just the Gaussian type and it is easy to
perform. 
Similarly, the $v$-integration is well-approximated by the Gaussian in the
small $\beta$ limit:
\begin{align}
\int_{-\frac{\pi}{\beta}}^{\frac{\pi}{\beta}} \! d v \ \exp
\left[
- \beta |\xi| (v \pm v_0)^2
\right] \simeq 
\int_{-\infty}^{\infty} \! d v \ \exp
\left[
- \beta |\xi| (v \pm v_0)^2
\right] = \sqrt{\frac{\pi}{\beta |\xi|}}.
\label{eq:tint}
\end{align}
Therefore we find the Dirac zero-mode $\Psi$ in \eqref{eq:psi_sol} is
correctly normalized:
\begin{align}
\int^{\infty}_{-\infty} \! du \
 \int_{-\frac{\pi}{\beta}}^{\frac{\pi}{\beta}} \! dv \ \Psi^{\dagger} \Psi 
\simeq 
\frac{\beta}{\pi} |\xi| \int^{\infty}_{-\infty} \! du \ \int^{\infty}_{-\infty}
 \! dv \ 
\left[
\begin{array}{cc}
|E_{-}|^2 & 0 \\
0 & |E_{+}|^2
\end{array}
\right]
= \mathbf{1}_2.
\end{align}
Now we have found the Dirac zero-mode \eqref{eq:psi_sol}.
Through the Nahm transformation \eqref{eq:Nahm_transformation}, we are going to write down the
solution for the physical fields.
In the following, we derive the explicit monopole solution to the
equation \eqref{eq:monopole}.

\subsection{Adjoint scalar field}
The solution to the adjoint scalar field is obtained as  
\begin{align}
\Phi =& \ \im \int^{\infty}_{-\infty} \! u du \ \int_{-\frac{\pi}{\beta}}^{\frac{\pi}{\beta}} \! dv \
 \Psi^{\dagger} \Psi 
\simeq 
 - \im u_0 \tau_3 = -\im \text{Re} (s_0) \tau_3,
\label{eq:adjoint_higgs}
\end{align}
where we have approximated the $v$-integration by the Gaussian in the
small $\beta$ limit.
As we have indicated, at the zero $s_0 = 0$ 
, where $x = \pm C, y = 0$
, we have $r_0 =0$ and $\Phi$ becomes trivial.
Using the explicit form of $u_0$ given in \eqref{eq:rt_def},
in the asymptotic region $x,y \gg C$, the solution
\eqref{eq:adjoint_higgs} behaves like 
\begin{align}
\Phi \sim \text{const.} - \frac{\im}{\beta} \log \left(\frac{r}{C}\right) \tau_3
\end{align}
where $r^2 = x^2 + y^2$.
This is the desired asymptotic behavior of the monopole \eqref{eq:bc}.
Note that this $\tau_3$ 
represents a $U(1)$ Cartan subgroup of $SU(2)$.
A gauge invariant quantity $\mathrm{Tr} \Phi^2$ is evaluated as
\begin{align}
\mathrm{Tr} [\Phi^2] = - 2 u_0^2.
\label{eq:trp2}
\end{align}
One observes that this is completely $z$-independent which implies that the
solution represents a codimension two object.

\subsection{Gauge field}
We proceed to construct the gauge field configuration.
It is convenient to combine the gauge field as the $\zeta$ and
$\bar{\zeta}$ components. 
Then, the Nahm transformation becomes
\begin{align}
A_{\zeta} =& \ \frac{1}{2} (A_x - \im A_y) = \int^{\infty}_{-\infty} \!
 du \ \int_{-\frac{\pi}{\beta}}^{\frac{\pi}{\beta}} \! dv \ \Psi^{\dagger} \partial_{\zeta} \Psi, 
\notag \\
A_{\bar{\zeta}} =& \ \frac{1}{2} (A_x + \im A_y) = \int^{\infty}_{-\infty} \!
 du \ \int_{-\frac{\pi}{\beta}}^{\frac{\pi}{\beta}} \! dv \ \Psi^{\dagger} \partial_{\bar{\zeta}} \Psi, 
\notag \\
A_z =& \ \int^{\infty}_{-\infty} \!
 du \ \int_{-\frac{\pi}{\beta}}^{\frac{\pi}{\beta}} \! dv \ \Psi^{\dagger} \partial_z \Psi.
\end{align}
Here $\partial_{\zeta} = \frac{1}{2} (\partial_x - \im \partial_y)$,
$\partial_{\bar{\zeta}} = \frac{1}{2} (\partial_x + \im \partial_y)$.

We first evaluate the $u,v$-integrations in $A_z$.
Since the $z$-dependence is only inside the $E_{\pm}$, we have
\begin{align}
\Psi^{\dagger} \partial_z \Psi 
= - \frac{\im \beta |\xi|}{\pi}
\left[
\begin{array}{cc}
(v - v_0) e^{-\beta |\xi| \{(u-u_0)^2 + (v-v_0)^2 \}} 
& 
0 \\
0 
& 
(v + v_0) e^{- \beta |\xi| \{(u+u_0)^2 + (v+v_0)^2 \}} 
\end{array}
\right].
\end{align}
Again, in the small-$\beta$ limit, 
the $u,v$-integrals of the Nahm transformation is
approximated by the Gaussian and variants of it. 
Then we find the result is 
\begin{align}
A_z = 0.
\end{align}

Next, we calculate $A_\zeta$.
After  tedious calculations, 
we have 
\begin{align}
\Psi^{\dagger} \partial_{\zeta} \Psi = 
\frac{\beta}{2 \pi \sqrt{|\xi|}} 
\left[
\begin{array}{cc}
(\bar{\xi} \psi_{11} + |\xi| \psi_{21}) |E_{-}|^2 & (\bar{\xi} \psi_{12}
 + |\xi| \psi_{22}) E^{*}_{-} E_{+} \\
(- \bar{\xi} \psi_{11} + |\xi| \psi_{21}) E_{+}^{*} E_{-} & (- \bar{\xi}
 \psi_{12} + |\xi| \psi_{22}) |E_{+}|^2
\end{array}
\right]. 
\end{align}
Here the terms in each component are defined by 
\begin{align}
\psi_{11} 
=& \ \frac{|\xi|^{-\frac{1}{2}}}{\sqrt{\zeta^2 - C^2}} 
\left[
\frac{3}{4} C \cosh (\beta s_0) + \xi
\left\{
- \frac{\beta C}{4 |\xi|} \bar{\xi} |s-s_0|^2 \cosh (\beta s_0) +
 \frac{1}{\beta} (\frac{\beta}{2} |\xi| (\bar{s} - \bar{s}_0) + \frac{z}{2})
\right\}
\right],
\notag \\
\psi_{12} 
=& \  \frac{|\xi|^{-\frac{1}{2}}}{\sqrt{\zeta^2 - C^2}}
\left[
- \frac{3}{4} C \cosh (\beta s_0) + \xi 
\left\{
 \frac{\beta C}{4 |\xi|} \bar{\xi} |s + s_0|^2 \cosh (\beta s_0) +
 \frac{1}{\beta} (\frac{\beta}{2} 
 |\xi| (\bar{s} + \bar{s}_0) + \frac{z}{2})
\right\}
\right],
\notag \\
\psi_{21} 
=& \ \frac{|\xi|^{-\frac{1}{2}}}{\sqrt{\zeta^2 - C^2}}
\left[
\frac{\bar{\xi} C}{4 |\xi|} \cosh (\beta s_0)
+ |\xi| 
\left\{
- \frac{\beta}{2} |s-s_0|^2 \frac{\bar{\xi} C}{2 |\xi|} \cosh (2\pi s_0) +
 \frac{1}{\beta} (\frac{\beta}{2} |\xi| (\bar{s} - \bar{s}_0) + \frac{z}{2})
\right\}
\right],
\notag \\
\psi_{22} 
=& \ \frac{|\xi|^{-\frac{1}{2}}}{\sqrt{\zeta^2 - C^2}}
\left[
\frac{\bar{\xi} C}{4 |\xi|} \cosh (\beta s_0) - |\xi| 
\left\{
\frac{\beta}{2} |s+s_0|^2 \frac{\bar{\xi} C}{2 |\xi|} \cosh (\beta s_0) +
 \frac{1}{\beta} (\frac{\beta}{2} |\xi| (\bar{s} + \bar{s}_0) + \frac{z}{2})
\right\}
\right].
\end{align}
Finally, we perform the integrations over $u,v$. 
Again, the integrations are approximated by the Gaussian or its variants
in the small-$\beta$ limit.
After calculations, we find
\begin{align}
A_{\zeta} (x,y,z) 
= 
\frac{1}{2 \sqrt{\zeta^2 - C^2}}
\left(
\begin{array}{cc}
\frac{\zeta}{2 \xi} + \frac{z}{\beta} & 
- \frac{\zeta}{2 \xi} e^{-2\im v_0 z} e^{-\beta |\xi|
|s_0|^2} \\
- \frac{\zeta}{2 \xi} e^{2 \im v_0 z} e^{- \beta |\xi| |s_0|^2} 
& \frac{\zeta}{2 \xi} - \frac{z}{\beta}
\end{array}
\right).
\label{eq:gauge_zeta}
\end{align}
One notices that this expression does not exhibit the traceless
condition of the $SU(2)$ algebra.
However, we observe that the expansion of the following quantity 
in the $C \gg x,y$ region,
\begin{align}
\cosh^{-1} \frac{1}{2} 
\left(
\sqrt{
\left(
1 + \frac{x}{C}
\right)^2 + 
\left(
\frac{y}{C}
\right)^2
}
+
\sqrt{
\left(
1 - \frac{x}{C}
\right)^2 + 
\left(
\frac{y}{C}
\right)^2
}
\right)
=& \  \frac{y}{C} - \frac{1}{6} 
\left(
\frac{y}{C}
\right)^3 + \frac{1}{2} \left(\frac{y}{C}\right)
 \left(\frac{x}{C}\right)^2
+ \cdots.
\label{eq:arccosh_exp}
\end{align}
exhibits the leading order behavior of each quantity in the expression
\eqref{eq:gauge_zeta} in the large-$C$.
Namely, using the expression \eqref{eq:rt_def}, we have 
\begin{align}
u_0 \sim \mathcal{O} \left(\frac{y}{C} \right) 
, \qquad 
v_0 \sim \frac{\pi}{2 \beta} + \mathcal{O} \left( \frac{x}{C} \right)
, \qquad 
|s_0|^2 = u_0^2 + v_0^2 \sim \frac{\pi^2}{4 \beta^2} + 
\mathcal{O} 
\left(
\left(
\frac{x}{C}
\right)^2, 
\left(
\frac{y}{C}
\right)^2
\right).
\end{align}
Here we have chosen the $n=0$ branch in the definition of $v_0$.
Then, we find 
\begin{align}
\beta |\xi| |s_0|^2 
\sim \frac{C}{\beta} + 
\mathcal{O} 
\left(
\left(
\frac{x}{C}
\right)^2, 
\left(
\frac{y}{C}
\right)^2
\right).
\end{align}
Therefore, the off-diagonal parts in \eqref{eq:gauge_zeta} 
behave like $\sim e^{- \frac{C}{\beta}}$ and they are exponentially
suppressed and ignored compared to the diagonal part in the large-$C$
(or small-$\beta$) limit.
Furthermore, the term $\frac{\zeta}{2\xi}$ in the diagonal part is small
compared to $\frac{z}{\beta}$ and suppressed over a $\mathcal{O}(\beta/C)$ quantity.
Therefore, we find the gauge field solution in the large-$C$ 
(and small-$\beta$) limit is
\begin{align}
A_{\zeta} \sim \frac{z}{2 \beta \sqrt{\zeta^2 - C^2}}
\left(
\begin{array}{cc}
1 & 0 \\
0 & -1
\end{array}
\right).
\label{eq:gauge_zeta_large-C}
\end{align}
This expression satisfies the traceless condition of the $SU(2)$ algebra
as expected. 
In summary, we obtain the gauge field of the codimension two monopole as 
\begin{align}
A_x = 
\frac{z}{2 \beta} 
\left(
\frac{1}{\sqrt{\zeta^2 - C^2}} - \frac{1}{\sqrt{\bar{\zeta}^2 - C^2}}
\right) \tau_3, \quad 
A_y =
\frac{z}{2 \beta} 
\left(
\frac{1}{\sqrt{\zeta^2 - C^2}} + \frac{1}{\sqrt{\bar{\zeta}^2 - C^2}}
\right) \tau_3, \quad 
A_z =
0.
\label{eq:gauge_field_solution}
\end{align}
Surprisingly, the monopole solution \eqref{eq:adjoint_higgs}, \eqref{eq:gauge_field_solution}
we have obtained in the large-$C$ limit via the Nahm construction is an
exact solution to the BPS equation \eqref{eq:monopole}.
One can easily confirm that the solution \eqref{eq:adjoint_higgs},
\eqref{eq:gauge_field_solution} satisfies the equation \eqref{eq:monopole} for any values of $C$.
We will comment on this issue later.

Since the gauge invariant quantity \eqref{eq:trp2} is independent of $z$, the
$z$-dependence in \eqref{eq:gauge_field_solution} is just due to the gauge artifact.
In order to see this fact explicitly, we express the leading order form
of the solution in $\mathcal{O} (x^i/C)$.
First, at large-$C$, we find the approximated solution is 
\begin{align}
A_x \simeq \frac{\im z}{\beta C} \tau_3, \quad 
A_y \simeq 0, \quad 
A_z \simeq 0.
\end{align}
From these we find
\begin{align}
B_x \simeq 0, \quad
B_y \simeq \frac{\im}{\beta C} \tau_3, \quad 
B_z \simeq 0.
\end{align}
Next, using the expansion of \eqref{eq:arccosh_exp}, 
we find the leading order behavior of the adjoint
scalar field is
\begin{align}
\Phi = - \im u_0 \tau_3 
\simeq
- \im \frac{y}{\beta C}
\tau_3 + \mathcal{O} ((x^i)^3/C^3).
\end{align}
One confirms that these expressions indeed satisfy the BPS equation \eqref{eq:monopole}.
It is now straightforward to find a gauge transformation that makes the
solution be $z$-independent.
The gauge transformation 
\begin{align}
\Phi \to \Phi' = U \Phi U^{\dagger}, \quad 
A_i \to A_i' = U A_i U^{\dagger} + U \partial_i U^{\dagger}, \quad 
U \in SU(2).
\end{align} 
with 
\begin{align}
U = \mathbf{1}_2 + \im \frac{xz}{\beta C} \tau_3, \quad 
U^{\dagger} = \mathbf{1}_2 - \im \frac{xz}{\beta C} \tau_3,
\end{align}
makes it possible to remove the $z$-dependence. 
The result is 
\begin{align}
A'_x \sim 0, \quad A'_y \sim 0, \quad A'_z \sim - \frac{\im x}{\beta C}
 \tau_3, \quad \Phi' \sim - \frac{\im y}{\beta C} \tau_3.
\label{eq:z-indep_sol}
\end{align}
Therefore the solution represents completely codimension two object.
This is similar to the situation where the 't Hooft monopole of
codimension three is obtained by the periodic instanton on $\mathbb{R}^3 \times
S^1$ \cite{Rossi:1978qe, Harrington:1978ve}.
In there the $S^1$ dependence of the periodic instanton solution 
is completely gauged away and the resulting solution is independent of
the periodic direction.

\section{Heterotic gauge five-brane with monopole of codimension two}
In this section, based on the monopole of codimension two discussed in the previous
section, we construct the gauge five-brane solution in heterotic
supergravity.
In the ansatz \eqref{eq:five-branes}, we compactify the transverse
directions $x^3$ and $x^4$ to $T^2 = S^1 \times S^1$ and consider the
small $T^2$ limit.
The self-duality equation for the gauge field effectively reduces to
that in two dimensions $\mathbb{R}^2$.
As a solution to this equation, we employ the small $S^1$ limit of the
monopole solution in $\mathbb{R}^2 \times S^1$.
The solution is given by \eqref{eq:adjoint_higgs} and \eqref{eq:gauge_field_solution}.
In particular, we identity the adjoint scalar field $\Phi$ with $A_4$ component. 
The Poisson equation \eqref{eq:source} reduces to  
\begin{align}
\partial_i^2 H = 
4 \alpha' \partial_i \mathrm{Tr} [B_i \Phi]
+ \mathcal{O}(\alpha^{\prime 2}), \quad  (i=1,2,3),
\end{align}
where the source term in the right hand side 
is provided by the one for the monopole of codimension two.
Using the monopole equation \eqref{eq:monopole} and the solution
\eqref{eq:trp2}, the first term in the right hand side is rewritten as 
\begin{align}
4 \alpha' \partial_i \mathrm{Tr} [B_i \Phi] = - 2 \alpha' 
 \partial_i^2 \mathrm{Tr} [\Phi^2] =  4 \alpha' \partial_i^2 u_0^2.
\end{align}
Therefore, the $H$-function is determined to be
\begin{align}
H (x,y) = h_0 + \frac{4 \alpha'}{\beta^2} 
\left[
\cosh^{-1} \frac{1}{2C} (\sqrt{(x+C)^2 +y^2} + \sqrt{(x - C)^2 +y^2})
\right]^2.
\label{eq:sol_H}
\end{align}
Here $h_0$ is a constant.
The metric, dilaton and the NS-NS $B$-field are determined through the BPS
ansatz \eqref{eq:five-branes}.
A particular emphasis is placed on the fact that the dilaton field 
$e^{2 \phi} = H(x,y)$ never becomes imaginary valued when $h_0 \ge 0$.
This is contrasted with the gauge solution based on the smeared monopole
\cite{Sasaki:2016hpp}. 
Even more, the metric governed by the $H$-function 
 \eqref{eq:sol_H}
is well-defined in $\mathbb{R}^2$.
The asymptotic behavior of the harmonic function looks like
\begin{align}
H (r) \sim \frac{4 \alpha'}{\beta^2} [\log \left( \frac{r}{C} \right)]^2,
 \qquad (r \to \infty).
\label{eq:asympt_gauge}
\end{align}
This is compared with the smeared solution $H (r) \sim h_0 - \frac{2\alpha'}{r^2
[\log (r/\mu)]^2} \ (r \to \infty)$.
We note that the asymptotic behavior \eqref{eq:asympt_gauge} of the
gauge five-brane is quite different from that of 
an authentic harmonic function $H (r) \sim \log r \ (r \to \infty)$ for
codimension two branes in type II theories.

The Ricci scalar for the geometry is calculated to be 
\begin{align}
R (\omega) =& \  - \frac{ 24 h_0 \alpha' \beta^4}
{
\sqrt{(x - C)^2 + y^2}
\sqrt{(x + C)^2 + y^2}
}
\notag \\
& \ \times 
\left(
h_0 \beta^2 + 4 \alpha'
\text{arccosh}^2
\frac{1}{2C}
\left(
\sqrt{(x - C)^2 + y^2}
+
\sqrt{(x + C)^2 + y^2}
\right)
\right)^{-3}.
\end{align}
Here the definition of the Ricci scalar is given in \eqref{eq:Ricci}.
One observes that the Ricci curvature of the geometry asymptotically
vanishes.
Remarkably, we find that all the components of the Ricci tensor $R_{MN}$
vanishes at the infinity of $\mathbb{R}^2$. 
Therefore, the geometry is asymptotically Ricci flat. 
This is in contrast to the codimension two stand-alone objects in 
type II string theories \cite{Greene:1989ya, Gibbons:1995vg, Bergshoeff:2011se}. 
Indeed, any supergravity solutions for stand-alone codimension two
objects obtained so far have only their description near the core of
branes \cite{Kikuchi:2012za}. 

\begin{figure}[tb]
\centering
\includegraphics[scale=0.8]{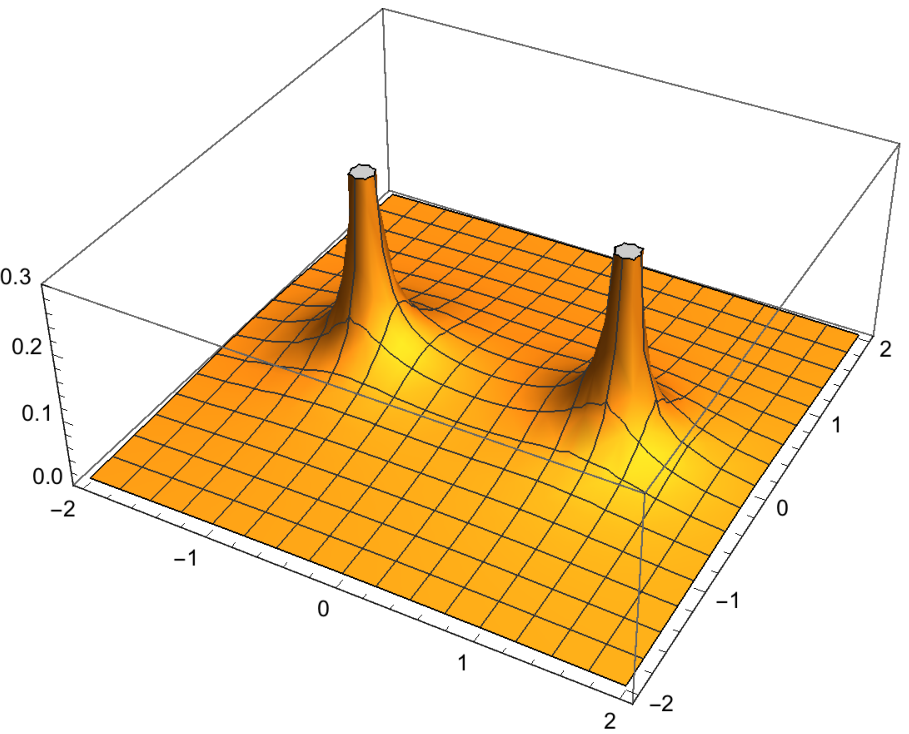}
\includegraphics[scale=0.7]{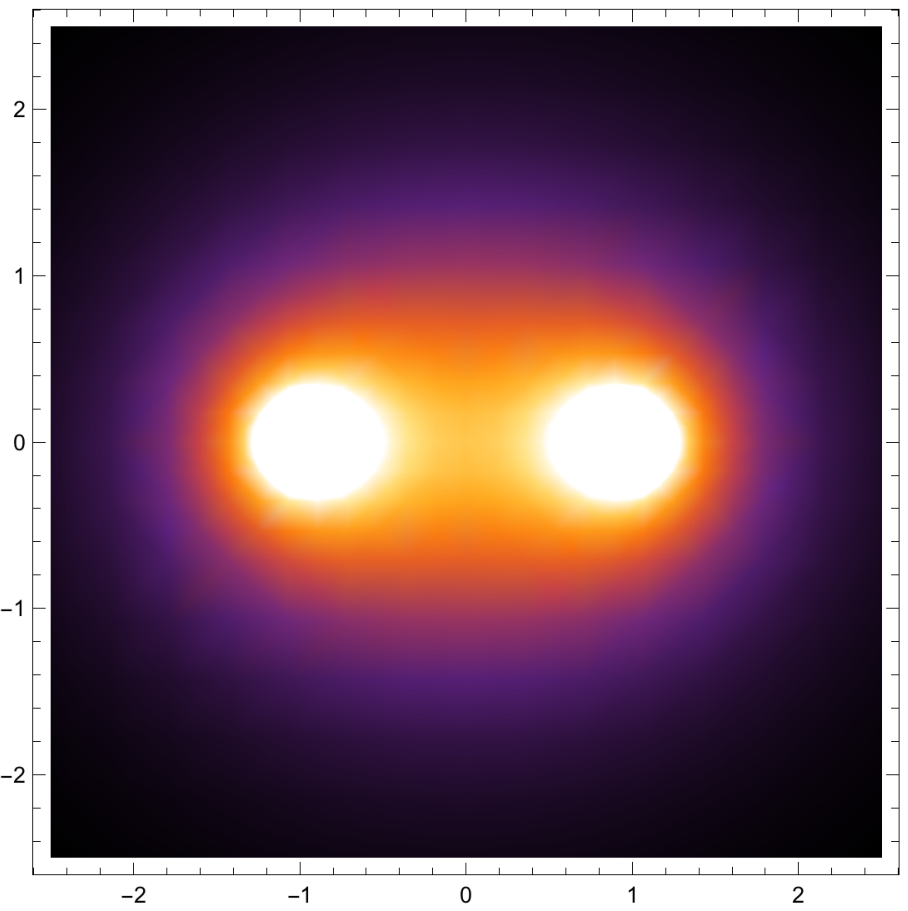}
\caption{
The energy (topological charge) density 
$\frac{1}{4} \text{Tr} [F_{mn} \tilde{F}_{mn}] = \partial_i^2 v_0^2$
for the monopole in the $(x^1,x^2)$ plane (left).
The parameters are $C=1$, $\beta = 1$.
The absolute value of the Ricci scalar in the $(x^1,x^2)$ plane (right).
}
\label{fig:monopole_energy}
\end{figure}

The plots of the energy density for the monopole 
and the absolute value of the Ricci scalar are found in fig.\ref{fig:monopole_energy}.
One finds that there are two poles in the energy density and the scalar
curvatures where the quantities diverge.
The origin of these two poles is obvious from the viewpoint of the Nahm
construction in compact spaces.
As is evident from the expression \eqref{eq:Dirac_operator}, 
the Hitchin field solution $P (s)$ on the cylinder $\mathbb{R} \times S^1$
clearly breaks the spherical symmetry in the physical space even for the
single monopole case. 
This substantially leads to the axially symmetric monopole solution
\eqref{eq:adjoint_higgs}, \eqref{eq:gauge_field_solution}.
This breakdown of spherical symmetry seems a fate of codimension two
monopoles \cite{Maldonado}.

As we have mentioned before, the solution (\ref{eq:adjoint_higgs}) and
(\ref{eq:gauge_field_solution}) is valid for any values of $C$.
The parameter $C$ corresponds to the distance between two centers of the
energy peaks and the supergravity solution becomes axially symmetric due to the existence
of the $C$-parameter.
This symmetry is quite different from the known five-brane solutions which have spherical symmetry. 
It is worthwhile to 
examine the $C\rightarrow0$ limit of the solution and look
for a spherically symmetric solution. 
It is easy to take a $C\rightarrow0$ limit for the gauge fields.
However, it is not straightforward to consider the limit $C \to 0$
for the adjoint scalar field.
We evaluate only a dominant term for the adjoint scalar in the $C\rightarrow0$ limit:
\begin{align}
\Phi|_{ \tiny C \rightarrow 0 }  &= \lim_{C\rightarrow0} { \im \over \beta } \cosh^{-1} { \sqrt{x^2+y^2} \over C }  \tau_3 
, \nonumber\\
A_x|_{ \tiny C \rightarrow 0 } &= { z \over 2 \beta } \left( { 1 \over \zeta } - { 1 \over \overline{\zeta} } \right) \tau_3 
,~~
A_y|_{ \tiny C \rightarrow 0 } = {  z \over 2 \beta } \left( { 1 \over \zeta } + { 1 \over \overline{\zeta} } \right) \tau_3 
,~~
A_z|_{ \tiny C \rightarrow 0 } = 0 
, \label{Czerolimit}
\end{align}
Using this expression, $D_i\Phi$ is calculated as
\begin{align}
D_1 \Phi|_{ \tiny C \rightarrow 0 } &= \lim_{C\rightarrow0}{ \im x  \over \beta \sqrt{ x^2 + y^2 } \sqrt{(x^2+y^2) - C^2}} \tau_3 = { \im x \over \beta (x^2+y^2) } \tau_3, \nonumber\\
D_2 \Phi|_{ \tiny C \rightarrow 0 } &= \lim_{C\rightarrow0}{ \im y  \over \beta \sqrt{ x^2 + y^2 } \sqrt{(x^2+y^2) - C^2}} \tau_3 = { \im y \over \beta (x^2+y^2) } \tau_3, \nonumber\\
D_3 \Phi|_{ \tiny C \rightarrow 0 } &= 0.
\label{eq:C0_sol1}
\end{align}
We also obtain
\begin{align}
B_1 &= - { \im x \over \beta (x^2+y^2) }
\tau_3
,~~~~~B_2 =- { \im y \over \beta (x^2+y^2) }
\tau_3
,~~~~~B_3 =0.
\label{eq:C0_sol2}
\end{align}
Again, we confirm that the $C \to 0$ expressions \eqref{eq:C0_sol1} and
\eqref{eq:C0_sol2} indeed satisfy the BPS monopole equation
(\ref{eq:monopole}).
We then find a heterotic five-brane solution based on  (\ref{eq:five-branes}) as
\begin{align}
H|_{ \tiny C \rightarrow 0 }&= h_0 + \lim_{C\rightarrow0}{ 4 \alpha^{\prime} \over \beta^2 } \left[ \cosh^{-1} { \sqrt{x^2+y^2} \over C } \right]^2, ~~~~e^{2\phi|_{ \tiny C \rightarrow 0 }} = H|_{ \tiny C \rightarrow 0 }, \nonumber\\
g_{\mu\nu}|_{ \tiny C \rightarrow 0 } &= \eta_{\mu\nu},~~~~ g_{mn}|_{ \tiny C \rightarrow 0 } = e^{2\phi|_{ \tiny C \rightarrow 0 }} \delta_{mn}, ~~~~\hat{H}_{mnp}|_{ \tiny C \rightarrow 0 }= -{1\over2} 
\varepsilon_{mnpq} 
\partial_q H|_{ \tiny C \rightarrow 0 },\nonumber\\
A_x|_{ \tiny C \rightarrow 0 } &= { z \over 2 \beta } \left( { 1 \over \zeta } - { 1 \over \overline{\zeta} } \right) \tau_3,~~
A_y|_{ \tiny C \rightarrow 0 } = {  z \over 2 \beta } \left( { 1 \over \zeta } + { 1 \over \overline{\zeta} } \right) \tau_3,~~
A_z|_{ \tiny C \rightarrow 0 } = 0, \nonumber\\
A_4 |_{ \tiny C \rightarrow 0 } &= \lim_{C\rightarrow0} { \im \over \beta } \cosh^{-1} { \sqrt{x^2+y^2} \over C } \tau_3,~~~~A_{\mu}|_{ \tiny C \rightarrow 0 } =0. 
\label{C0hetero}
\end{align}
Under the $C\rightarrow0$ limit, we find that the Ricci tensor converges to zero. 
This means that the spherical symmetric solution becomes globally Ricci flat. 

\section{Kaluza-Klein gauge five-brane and gauge $5^2_2$-brane}
In this section, we perform the T-duality transformations along the two
isometries on the solution found in the previous section.
We will write down the regular solutions for the KK5- and $5^2_2$-branes
associated with the gauge five-brane.
Specifically, we are interested in the $z$-independent solution
obtained in \eqref{eq:z-indep_sol}
:
\begin{align}
H&= h_0 + { 4 \alpha^{\prime} \over \beta^2 C^2 } y^2,~~
e^{2\phi} = H, ~~
g_{mn} = e^{2\phi} \delta_{mn} ,~~ 
g_{\mu\nu} =\eta_{\mu\nu},~~ \nonumber\\
B_{34}&= -\alpha^{\prime} { 4 xy \over \beta^2 C^2 },~~~
A_3= \im { x \over \beta C} \tau_3
,~~~
A_4=-\im { y \over \beta C } \tau_3
, 
\label{H f solution}
\end{align}
The other components are zero.
We note that this solution is available up to $\mathcal{O} ((x^i)^3/C^3)$.

For the heterotic supergravity action in the first order in
$\alpha^{\prime}$, 
the T-duality transformation rule, called 
heterotic Buscher rule, is written as \cite{Sasaki:2016hpp, Hohm:2014sxa}, 
\begin{align}
G_{MN} &= g_{MN} - B_{MN} + 2 \alpha^{\prime} \tr A_M A_N, \nonumber\\
\tilde{g}_{MN} &= g_{MN} + {1 \over G^{2}_{nn}} ( g_{nn} G_{nM} G_{nN} -G_{nn} g_{nM} G_{nN} -G_{nn} G_{nM} g_{nM} ), \nonumber\\
\tilde{B}_{MN} &= B_{MN} + {1\over G_{nn}} (G_{nM}B_{Nn} - G_{nN} B_{Mn}),~~
\tilde{g}_{nM} = -{ g_{nM} \over G_{nn} } + { g_{nn} G_{nM} \over G^{2}_{nn} },~~
\tilde{g}_{nn} = - { g_{nn} \over G^{2}_{nn} }, \nonumber\\
\tilde{B}_{nM} &= - { 1 \over G_{nn}} ( B_{nM} + G_{nM} ),~~
\tilde{\phi} = \phi - {1\over2} \log|G_{nn}|,~~
\tilde{A}^I_{n} = - { A^I_n \over G_{nn} },~~
\tilde{A}^I_{M} = A^I_M - {G_{nM} \over G_{nn}} A^I_n,
\end{align}
where the index ``$n$'' means the T-dualized (isometry) direction and the tilde represents dualized fields. 
The combination 
\footnote{
Here, we have chosen the combination $g_{MN} - B_{MN}$ not $g_{MN} + B_{MN}$ 
which may be widely used in literature.
This choice originates from 
the convention related to the generalized spin connection $\omega_{\pm}$.
Since we have defined the $R^2$ term in the action by $\omega_+$, we
have to choose the abovementioned combination in $G_{MN}$.
Otherwise the $O(d,d)$ T-duality symmetry is not realized.
}
$g_{MN} - B_{MN}$ is a primitive metric in
the double field theory (DFT) \cite{Hull:2009mi}.

\begin{figure}[t]
\centering
\includegraphics[scale=0.8]{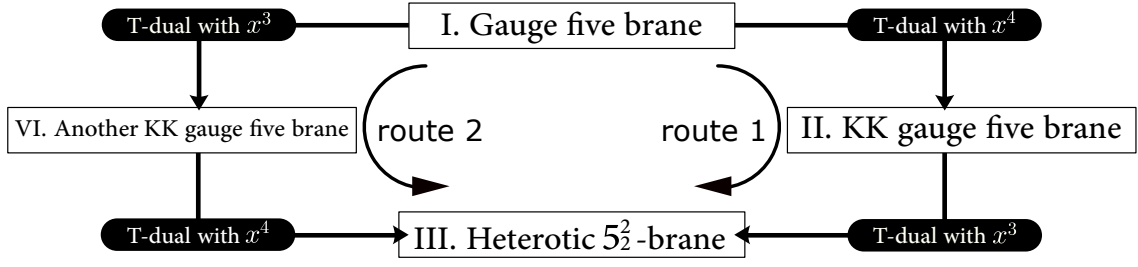}
\caption{
The picture indicates the T-duality relations 
among heterotic five-branes. 
Since the gauge five brane has two isometry directions $(x^3,x^4)$, there are two routes to obtain heterotic $5^2_2$-brane. 
}
\label{T-dualityR}
\end{figure}

Since the fields (\ref{H f solution}) have two isometry directions, there are two routes to obtain the $5^2_2$-brane. 
We follow the route 1 in fig. \ref{T-dualityR} and show the Kaluza-Klein
gauge five-brane and $5^2_2$-brane in this section.
For another Kaluza-Klein gauge five-brane, we show it in the Appendix in detail. 
\subsection{Kaluza-Klein gauge five-brane}
First, we perform a T-duality transformation with the gauge solution and
obtain a T-dualized object 
-- the KK5-brane of codimension two. 
When we take the heterotic T-duality with respect to the $x^4$-direction
for the fields (\ref{H f solution}), we obtain the Kaluza-Klein gauge
five-brane which corresponds to ``II. KK gauge five brane'' in fig. \ref{T-dualityR}:
\begin{align}
H &= h_0 + { 4 \alpha^{\prime} \over \beta^2 C^2 }y^2,~~e^{2\phi^{(4)}} =   { H \over h_0 }, \nonumber\\
g^{(4)}_{ab} &= g_{ab} = H \delta_{ab},~~g^{(4)}_{33} = H,~~g^{(4)}_{44}= {1\over h_0^2} H,~~g_{\mu\nu}^{(4)}=\eta_{\mu\nu}, \nonumber\\
B^{(4)}_{34} &= {\alpha^{\prime} \over h_0} { 4 xy \over \beta^2 C^2},~~
(A_3)^{(4)}=\im { x \over \beta C } \tau_3 
,~~~
(A_4)^{(4)}={ \im \over h_0 }  { y \over \beta C } \tau_3
, 
\label{KaluzaKlein}
\end{align}
where $(4)$ means the T-dualized direction and the indices $a,b=1,2$. 
It is obvious that the $B^{(4)}$ field has a non-zero component and the dilaton $e^{2\phi^{(4)}}$ is regular. 
Unlike the usual Kaluza-Klein five brane (Taub-NUT), the solution has
non-zero components of the $B$-field and the gauge field. 
Therefore, we can  regard the brane as a source of the $H$-flux.
We also confirm that the gauge field satisfies the self-duality
condition, 
$F_{mn} = { 1 \over 2 } \varepsilon_{mn}^{~~~pq}F_{pq} $ in two dimensions.
This sign flip originates from the convention of the heterotic Buscher
rule \cite{Sasaki:2016hpp}. 

The fields (\ref{KaluzaKlein}) are almost the same with (\ref{H f
solution}) except for the constant coefficient. 
The reason is that the component of the extended metric
$G_{44}$ becomes a constant, 
and somehow it is the same situation as in the smeared gauge solution
discussed in \cite{Sasaki:2016hpp, Callan:1991dj}. 

\subsection{Heterotic $5^2_2$-brane}
Now, we perform the second T-duality transformation along the $x^3$ direction
on the Kaluza-Klein solution. 
As a result, the solution obtained corresponds to the ``III. Heterotic
$5^2_2$-brane'' in fig.\ref{T-dualityR}:
\begin{align}
H &= h_0 + { 4 \alpha^{\prime} \over \beta^2 C^2 }y^2,~~e^{2\phi^{(43)}} = {1\over h_0^2} \Bigl( h_0 + { 4 \alpha^{\prime} \over \beta^2 C^2 } x^2  \Bigr) \nonumber\\
g^{(43)}_{ab} &=  H \delta_{ab},~~
g^{(43)}_{33} = {1\over h_0^2} \Bigl( h_0 + { 4 \alpha^{\prime} \over \beta^2 C^2 } \bigl(  2 x^2 - y^2 \bigr) \Bigr),~~
g^{(43)}_{34}=-8 { \alpha^{\prime} \over h_0^2 } { xy \over \beta^2 C^2 },~~
g^{(43)}_{44}= {H \over h_0^2},~~g_{\mu\nu}^{(43)}=\eta_{\mu\nu}, \nonumber\\
B^{(43)}_{34} &= 4 { \alpha^{\prime} \over h_0^2 } { xy \over \beta^2  C^2 },~~
(A_3)^{(43)}=-{ \im \over h_0 } { x \over \beta C } \tau_3
,~~
(A_4)^{(43)}={ \im \over h_0 }  { y \over \beta C } \tau_3
.  \label{x43T-dualized}
\end{align}
Here, the fields are written in ${\cal O}((x^i)^2/C^2)$. 
It is clear that the dilaton $\phi^{(43)}$ is regular and does not take
a negative value, unlike the one in \cite{Sasaki:2016hpp}. 

For this solution, the components of the $B$-field are non-zero.
One can easily write down the generalized metric \cite{Hohm:2014sxa} 
associated with the solution \eqref{x43T-dualized}. 
By the generalized metric, 
we find that the monodromy around the gauge $5^2_2$-brane
solution we obtained is trivial and it does not exhibit any non-geometric feature. 
In other words, the gauge $5^2_2$-brane in heterotic theories is not a non-geometric but a 
{\it geometric} object. 
This is in sharp contrast to the exotic $5^2_2$-branes of the neutral and
 symmetric types \cite{Sasaki:2016hpp} and those in type II theories \cite{deBoer:2010ud}.
We also find that the gauge field strength
 in (\ref{x43T-dualized}) satisfies the anti self-duality condition,
 $F_{\mu\nu} = - { 1 \over 2 } \varepsilon_{mn}^{~~~pq}F_{pq} $.


\section{Conclusion and discussions}
In this paper, we studied the BPS gauge five-brane solution of
codimension two in heterotic supergravity.
The 1/2 BPS ansatz reveals the fact that the $H$-function 
is determined through the source term given by the monopoles of codimension
two. The desired monopole is described by the small circle limit of
the monopole chain defined in $\mathbb{R}^2 \times S^1$. 
An analytic solution of the monopole chain is explicitly written down by the
Nahm construction discussed in \cite{Ward:2005nn}.
Using this solution, we find the explicit form of the gauge five-brane
of codimension two.
We found two poles in the curvature of the geometry for the gauge five-brane. 
This is due to the fact that the monopole we constructed preserves only
the axial symmetry in $\mathbb{R}^2$.
This may be interpreted from the viewpoint of solitons in compact spaces.
The multipole structure of solutions is quite common in the case of periodic instantons where these
multipoles correspond to monopole constituents of a single instanton
\cite{Kraan:1998sn}.
It is well known that solitons in compact spaces with non-trivial
asymptotic holonomy possesses constituents inside each energy peak.
In the analysis of the monopole, we employed the boundary condition
where the solution behaves like the $U(1)$ Dirac monopole at
asymptotics. In other words, the $SU(2)$ gauge symmetry is broken down
to $U(1)$ at infinity.
This breaking is characterized by the asymptotic holonomy.
As discussed in \cite{Maldonado}, for the monopole in two dimensions, 
there is always a non-trivial asymptotic holonomy
due to the logarithmic growth of the adjoint scalar fields.
This substantially leads to the introduction of constituents for the
monopoles.
This phenomenon is interpreted as D-brane configurations \cite{Lee:1997vp} in type II theories.
It is interesting to explore whether the same
kind of interpretation is possible in heterotic theories.

The parameter $C$ in the solution controls the breaking of the spherical
symmetry. We made an analysis on the $C \to 0$ limit of the solution
where the spherical symmetry is expected to be realized. 
We also found that the asymptotic geometry is Ricci flat 
which is consistent with the fact that the monopole charge density becomes zero at the asymptotic region. 
Despite this fact, however, the total energy (or the topological charge) associated with the
monopole of codimension two diverges.
This is an inevitable fate of codimension two objects.

The $H$-function 
that governs the solution behaves like $H (x,y)
\sim [\log r]^2$ in the asymptotics which is contrasted with the gauge solution based on
the smeared monopole discussed in our previous paper \cite{Sasaki:2016hpp}.
The geometry based on the smeared monopole has been ill-defined in some regions in space-time 
and the dilaton becomes imaginary valued in there.
We stress that the new solution based on the monopole chain in this
paper overcome this problematic property.

Since the solution exhibits the $U(1)^2$ isometry along the transverse directions
to the brane world-volume, we can perform the chain of the T-duality transformations.
By applying the modified Buscher rule in heterotic theories, 
we performed the T-duality transformations of the gauge five-brane
of codimension two. We wrote down the KK gauge five-brane and the gauge
$5^2_2$-brane. The latter is a candidate of exotic branes in heterotic
string theories. We find that the monodromy of the gauge $5^2_2$-brane
is trivial and it is totally a geometric object. This is contrasted to the
symmetric and the neutral $5^2_2$-branes discussed in \cite{Sasaki:2016hpp}.

\begin{figure}[t]
\centering
\includegraphics[scale=0.9]{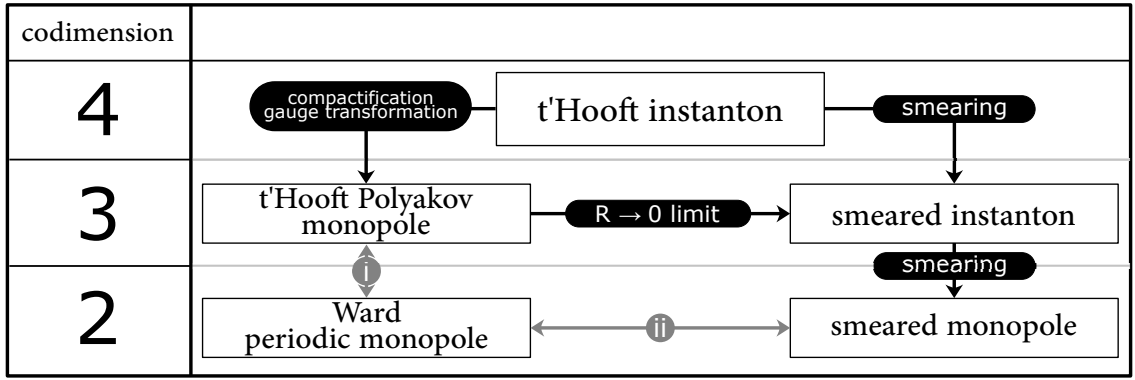}
\caption{
The relationships among five-branes of various codimensions.
}
\label{T-duality}
\end{figure}

It is also interesting to make contact with the heterotic gauge
five-branes of various codimensions (fig. \ref{T-duality}).
As we have discussed in the main body of this paper, 
the codimension three gauge five-brane based on the smeared instanton worked out in
\cite{Sasaki:2016hpp} just corresponds to the small circle limit of the
one found in \cite{Khuri:1992hk}.
There is an implication that the large circle limit of the Ward's
monopole chain becomes the ordinary 't Hooft monopole in three
dimensions \cite{Maldonado}.
We therefore expect that the codimension two five-brane we have obtained
reduces to the codimension three five-brane based on the 't Hooft
monopole in this limit (the relation (i) in fig.\ref{T-duality}).
On the other hand, we naively expect that a strict $\beta \to 0$ limit
of our solution results in the five-brane based on the smeared monopole
(the relation (ii) in fig.\ref{T-duality}).
However, it is difficult to observe this issue due to the lack of the
spherical symmetry of the Ward's solution.
It seems plausible that more subtle limit need to be considered as
in the case of the monopole limit of periodic instantons \cite{Gross:1980br}.

There are more interesting issues related to works done here and \cite{Sasaki:2016hpp}.
We have worked out the explicit T-duality chains of five-branes in
heterotic theories.
We have found that the three distinct five-branes exhibit totally
different behavior under the chain of the T-duality.
The $5^2_2$-brane of the symmetric and the neutral types are
non-geometric while the one of the gauge type is geometric.
What nature does it clarify this property?
The group theoretical classification of BPS multiplets in heterotic
theories based on Abelian gauge symmetries has been studied
\cite{Bergshoeff:2012jb}.
It would be interesting to study the structure of the BPS multiplet, 
especially the T-duality brane orbit \cite{Bergshoeff:2012ex} in
toroidary compactified heterotic theories.

It was also discussed that when the instanton 
in the gauge five-brane of codimension four shrinks to zero size $\rho \to
0$, a gauge multiplet on the brane world-volume becomes massless and the 
enhanced $SU(2)$ gauge symmetry is expected to appear \cite{Witten:1995gx}.
This small instanton limit of the gauge five-brane in the $SO(32)$
heterotic string theory is related to the D5-brane in type I theory via
the S-duality. 
The fact that the moduli spaces of periodic monopoles have hyperK\"ahler
metrics \cite{Cherkis:2000ft} together with the discussion \cite{Witten:1995gx} 
may lead to an interpretation of the gauge five-brane discussed in this
paper in type I theory side. 
Even more, it is interesting to study relations among various
five-branes in heterotic and type I theories. We will come back on these issues in the future.

\subsection*{Acknowledgments}
The authors would like to thank A.~Nakamula, J.~H.~Park and S.~J.~Rey  for useful discussions and comments. 
The work of S.~S. is supported in part by 
the Japan Society for the Promotion of Science (JSPS) KAKENHI Grant Number
JP17K14294 and Kitasato University Research Grant for Young Researchers.
The work of M.~Y.~ is supported by IBS-R018-D2.

\appendix
\section{Another Kaluza-Klein gauge five-brane}
Here we show another Kaluza-Klein gauge five-brane solution which
corresponds to ``IV. Another Kaluza-Klein gauge five-brane'' in fig. \ref{T-dualityR}.  
This is another configuration of the Kaluza-Klein gauge five-brane. 
Since the $(x,y)$ dependence of the fields in (\ref{H f solution}) is
asymmetric, the form of the T-dualized fields is slightly different from
the one in (\ref{KaluzaKlein}). 
The fields can be obtained by the heterotic T-duality along
$x^3$-direction for the solution (\ref{H f solution}): 
\begin{align}
H &= h_0 + { 4 \alpha^{\prime} \over \beta^2 C^2 }y^2,~~ e^{2\phi^{(3)}} = {1\over h_0} \Bigl( h_0 + { 4 \alpha^{\prime} \over \beta^2 C^2 } \Bigr), \nonumber\\
g^{(3)}_{ab} &=  H \delta_{ab},~~
g^{(3)}_{33} = {1\over h_0^2} \Bigl( h_0 + { 4 \alpha^{\prime} \over \beta^2 C^2 } \bigl( 2x^2 -y^2 \bigr) \Bigr),~~
g^{(3)}_{34}= 8 { \alpha^{\prime} \over h_0  } { xy \over \beta^2 C^2 },~~
g^{(3)}_{44}= {H},~~g_{\mu\nu}^{(3)}=\eta_{\mu\nu}, \nonumber\\
B^{(3)}_{34} &= -{4 \alpha^{\prime}\over h_0 } { xy \over \beta^2 C^2 },~~
(A_3)^{(3)}=-{ \im \over h_0 } { x \over \beta C } \tau_3
,~~
(A_4)^{(3)}=- \im   { y \over \beta C } \tau_3
. \label{x3T-dualized}
\end{align}
The fields are written in ${\cal O}((x^i)^2/C^2)$. 
There is a non-zero off-diagonal metric and the form of the dilaton
$\phi^{(3)}$ and a component of the metric $g^{(3)}_{33}$ are different
from (\ref{KaluzaKlein}). 
However, the gauge fields in (\ref{KaluzaKlein}) and (\ref{x3T-dualized}) are gauge equivalent. 
When we take a T-dual transformation along $x^4$ for
(\ref{x3T-dualized}), we can obtain the heterotic $5^2_2$-brane
(\ref{x43T-dualized}) as in fig. \ref{T-dualityR}.
Therefore, we confirm that the T-duality web for the heterotic gauge
five brane is closed.

\end{document}